\documentclass[fleqn,usenatbib]{mnras}

\usepackage{newtxtext,newtxmath}

\usepackage[T1]{fontenc}

\DeclareRobustCommand{\VAN}[3]{#2}
\let\VANthebibliography\thebibliography
\def\thebibliography{\DeclareRobustCommand{\VAN}[3]{##3}\VANthebibliography}

\usepackage{graphicx}	
\usepackage{amsmath}	

\usepackage[normalem]{ulem}
\usepackage{soul}

\newcommand{\Msun}{M_\odot}
\newcommand{\Mdot}{\dot{M}}
\newcommand{\Mdotstar}{\dot{M}_\ast}
\newcommand{\Mdotin}{\dot{M}_\mathrm{in}}

\newcommand{\Pdot}{\dot{P}}

\newcommand{\Edot}{\dot{E}}

\newcommand{\rin}{r_\mathrm{in}}
\newcommand{\rout}{r_\mathrm{out}}

\newcommand{\rco}{r_\mathrm{co}}
\newcommand{\rinmax}{r_\mathrm{in,max}}
\newcommand{\rA}{r_\mathrm{A}}
\newcommand{\reta}{r_\eta}

\newcommand{\Rinmax}{R_\mathrm{in,max}}
\newcommand{\DeltaR}{\Delta r/r_\mathrm{in}}

\newcommand{\Lacc}{L_\mathrm{acc}}
\newcommand{\Lcool}{L_\mathrm{cool}}
\newcommand{\Lx}{L_\mathrm{X}}

\newcommand{\Fx}{F_\mathrm{X}}
\newcommand{\Firr}{F_\mathrm{irr}}
\newcommand{\Tp}{T_\mathrm{p}}

\newcommand{\Teff}{T_\mathrm{eff}}
\newcommand{\cs}{c_{\rm s}}
\newcommand{\Md}{M_{\rm d}}

\newcommand{\Gammaacc}{\Gamma_\mathrm{acc}}
\newcommand{\GammaD}{\Gamma_\mathrm{D}}
\newcommand{\Gammadip}{\Gamma_\mathrm{dip}}

\newcommand{\gpers}{g~s$^{-1}$} 
\newcommand{\ergpers}{erg~s$^{-1}$} 
\newcommand{\spers}{s~s$^{-1}$}

\newcommand{\Alfven}{Alfv$\acute{\mathrm{e}}$n~}

\newcommand{\src}{1RXS J141256.0+792204}

\title[Evolution of Calvera-like CCOs]{Evolution of Calvera and Descendants of Calvera-like Central Compact Objects}

\author[A. A. Gen\c{c}ali et al.]{
A. A. Gen\c{c}ali,$^{1}$\thanks{E-mail: gencali@sabanciuniv.edu}
\"{U}. Ertan$^{1}$\thanks{E-mail: unal@sabanciuniv.edu},
F. Ertuğrul$^{1}$,
Ş. Demirok$^{1}$
and N. Niang$^{1}$
\\
$^{1}$Faculty of Engineering and Natural Sciences, Sabanc{\i} University, Orhanl{\i}, Tuzla, 34956, \.{I}stanbul, Turkey
}

\date{Accepted XXX. Received YYY; in original form ZZZ}

\pubyear{\the\year{}}

\begin{document}
\label{firstpage}
\pagerange{\pageref{firstpage}--\pageref{lastpage}}
\maketitle

\begin{abstract}
Calvera (\src) is an isolated neutron star recently classified as a central compact object (CCO) after its association with the supernova remnant (SNR) candidate G118.4+37.0. In this work, we investigate the long-term evolution and descendants of Calvera and the CCOs with similar initial conditions in the fallback disc model. We show that the observed spin period, period derivative, and X-ray luminosity of Calvera can be reproduced simultaneously with a magnetic field strength of $\simeq 4 \times 10^{10}$~G at the pole of the neutron star at an age of $\sim 10^4$~yr which is consistent with the estimated SNR age of the source. In the model, the lack of ordinary radio pulsations is due to ongoing mass accretion on to the star. From our simulations, we estimate that the source will become an ordinary radio pulsar (RP) after the termination of the accretion. The source will spin down under the weak dipole torque alone as an RP for a very long time ($\gtrsim 10^8$~yr) after the inactivation of the disc. From our simulations, we find that most of the CCOs with initial conditions similar to those of Calvera also become RPs which remain above or close to the upper border of the pulsar death valley after the inactivation of their discs. In the period–period derivative diagram, there is indeed a cluster of RPs in the region where the descendants of Calvera-like CCOs are estimated to be located in our model. 
\end{abstract} 

\begin{keywords}
accretion, accretion discs–stars: neutron–pulsars: individual: \src
\end{keywords}



\section{Introduction}

Central compact objects (CCOs) are isolated neutron stars (INSs) found near the centres of supernova remnants (SNRs). They are characterized by steady thermal X-ray emission with spectrum that can be fit by two blackbodies with temperatures in the range of $\sim 200-500$~eV with emitting radii ranging from $\sim 0.1$ to a few km. 
Their typical X-ray luminosities are in the $\Lx \sim 10^{32-34}$~\ergpers~range \citep{DeLuca2017}\footnote{\url{https://www.iasf-milano.inaf.it/~deluca/cco/main.htm}}.
They have no associated pulsar wind nebulae and could not be detected at optical, infrared, and radio wavelengths. 
Recently, very faint and highly variable pulsed radio emission was detected from a CCO, 1E 1207.4–5209 \citep{zhang2026}. No radio emission has been detected from the source in the archival data, which indicates that 1E 1207.4–5209 could be a transient radio source \citep{zhang2026}.
Among the $\sim 10$ confirmed CCOs, only three have measured spin periods, $P \sim 0.1 - 0.4$~s, and period derivatives, $\Pdot \sim (0.73 - 2.2) \times 10^{-17}$~\spers~\citep{Gotthelf2005,Halpern2010,Gotthelf2013CCOs}.
Their $\Lx$ is estimated to be one to two orders of magnitude higher than their rotational power, $\Edot = 4 \upi^2 I\Pdot / P^3 \sim 10^{32}$~\ergpers, where $I$ is the moment of inertia of the NS.
Their SNR ages are estimated to be in the range of $\sim 0.3 - 27$~kyr \citep{Gotthelf2013CCOs,DeLuca2017}.

1RXS J141256.0+792204 (also known as Calvera) was discovered by ROSAT as an X-ray source with a high X-ray-to-optical flux ratio \citep{Rutledge2008,Shevchuk2009}. The source is at a high Galactic latitude ($\sim 1.3 - 5$~kpc above the Galactic plane) with X-ray flux $\Fx \simeq 1.2 \times 10^{-12}$~\ergpers~cm$^{-2}$ ($0.1 - 2.4$~keV), and blackbody temperature $T_{\rm eff} = 215 \pm 25~$eV, which is a few times higher than that of known X-ray dim INSs (XDINs) but similar to that of CCOs \citep{Haberl2007,Rutledge2008,Zane2011}. There are uncertainties in its distance, and $\Lx$ is estimated as $\sim 10^{32}~d^2_{\rm kpc}$~\ergpers~\citep{Rutledge2008,Zane2011}. The best fit to its X-ray spectrum consistent with the pulse profile of the source is obtained for $d = 3.3$~kpc, assuming a NS radius of $13$~km \citep{Mereghetti2021}. For Calvera, $P$ and $\Pdot$ were measured to be $59.2$~ms and $3.3 \times 10 ^{-15}$~\spers~respectively, which corresponds to $\dot{E} \simeq 6.3 \times 10^{35}$~\ergpers~\citep{Zane2011, Halpern2013, Rigoselli2024}. Based on its $\Lx$, spectral, and rotational properties, Calvera could not be classified among the known INS populations. A candidate SNR, G118.4+37.0, was reported with the discovery of a ring of faint radio emission by LOFAR Two-metre Sky Survey in the neighbourhood of Calvera \citep{Arias2022}. Detection of $\gamma$-ray emission from the ring with the Fermi Large Area Telescope indicated that the source is associated with SNR G118.4+37.0, which placed the source into the CCO population \citep{Araya2023,Rigoselli2024,Daki2025}. The SNR age of the pulsar is estimated to be $< 10$~kyr \citep{Rigoselli2024}. From a detailed multi-wavelength analysis, the age of the SNR is estimated to be $\sim 10 - 20$~kyr corresponding to a distance of $\sim 4 - 5$~kpc \citep{Greco2025}. For this distance range, $\Lx \simeq (2.1 - 3.3)\times 10^{33}$~\ergpers.

In the models assuming CCOs are evolving in vacuum, the magnetic dipole field strength at the pole, $B_0$, is inferred from the dipole torque formula. For the three CCOs with measured $P$ and $\Pdot$, the estimated $B_0$ values are in the $(6 - 20) \times 10^{10}$~G range (on the equator the dipole field has half its polar value). 
In this dipole torque model, these $B_0$ values together with the measured periods are sufficient for pulsed radio emission \citep{Chen1993}. 
However, none of the CCOs have radio counterparts \citep{DeLuca2017}.
The lack of radio pulsations is generally attributed to the beaming geometry in this model \citep{Halpern2010}. 
If CCOs are not intrinsically radio silent, their immediate descendants could be found among ordinary radio pulsars (RPs). 
There are RPs with $B_0$ values, inferred from their $P$ and $\Pdot$, lower than and similar to those of the three CCOs. 
It was proposed that some of these sources could be disrupted recycled pulsars (DRPs) \citep{Lorimer2008}.
DRPs are estimated to be NSs that were mildly recycled through accretion in high-mass X-ray binaries (HMXBs), which were subsequently disrupted by a second supernova explosion \citep{Lorimer2004, Lorimer2008,Tauris2023}.
It was suggested that some of these RPs could be immediate descendants of CCOs instead of DRPs, 
and their cooling luminosity could be detected in X-rays \citep{Gotthelf2013DRP}.
To test this idea, they analysed $14$ RPs, but could not detect any CCO-like X-ray emission, which indicates that these RPs cannot be immediate descendants of CCOs \citep{Gotthelf2013DRP}.

Another mechanism proposed to explain the weak dipole fields and the long-term evolution of CCOs is the magnetic field burial scenario \citep{Ho2011,Vigano2013hiddenB}.
In this scenario, the magnetic field of sources is initially buried by the fallback material from the SN, and subsequently re-emerges, that is, the dipole field strength grows gradually in time \citep{Geppert1999, Geppert2013, Ho2011, Vigano2013hiddenB}.
For Calvera, \citet{Rigoselli2024} suggested that the field growth with a shorter diffusion time associated with a smaller fallback mass relative to those of the other pulsating CCOs could explain both its position in the $P-\Pdot$ diagram and its stronger dipole field.
In this case, the young descendants of CCOs could possibly be among the RPs with dipole fields stronger than those of known CCOs \citep{Gotthelf2013DRP}.
To test this idea, \citet{Luo2015} observed $12$ RPs with $B_0 \lesssim 2 \times 10^{11}$~G and could not detect any thermal X-ray radiation. 
They concluded that these sources are not likely to be immediate descendants of CCOs either \citep{Luo2015}.

In the fallback disc model, the torques acting on the NSs and the $B_0$ values estimated from their observed rotational properties are rather different from those estimated in the dipole torque model. Fallback discs could form around newborn NSs after supernova explosions \citep{Colgate1971,Michel1988,Chevalier1989,Perna2014}. It was suggested that NSs evolving with fallback disc can explain the observed $\Lx$ levels and $P$ clustering of anomalous X-ray pulsars (AXPs) \citep{Chatterjee2000}. \citet{Alpar2001} proposed that long-term evolutions of INSs with the initial conditions including the properties of fallback discs in addition to initial $P$ and $B_0$, could explain the emergence of not only AXPs, but also other INS populations during the long-term evolution. Fallback disc emission has been investigated extensively for different INS systems \citep{Perna2000,Ertan2006,Ertan2007,Ertan+2017,Posselt2018}. Characteristics of pulsed hard and soft X-ray radiation from AXPs were modelled by the spectral properties of emission from the accretion column and the poles of the NS \citep{Trumper2010,Trumper2013,Kylafis2014,Zesas2015}. The fallback disc model was developed by including the effects of X-ray irradiation of the disc, the contribution of the cooling luminosity to the irradiation flux, and the inactivation of the disc at low temperatures to investigate the long-term evolution of different INS populations \citep{Ertan2009,Ertan2014,Benli2016,Benli2017,Benli2018,BenliCCO2018,Gencali2018,Gencali2021,Gencali2022,Gencali2023}.
Based on the results of these earlier work, the evolutionary links between the INS populations were studied in detail by \citet{Gencali2024}.

In the fallback disc model, the $P$, $\Pdot$, and $\Lx$ values of CCOs can be reproduced with relatively low $B_0 \sim (2$–$5) \times 10^9$~G, while there is currently an ongoing mass accretion on to the magnetic poles of NS \citep{BenliCCO2018}. 
The radio emission is expected to be quenched by the mass-flow on to the NS, which is the reason for the lack of radio pulsations from CCOs in this model. 
After the termination of the accretion, they could become RPs provided that their rotational power is still sufficient for the pulsed radio emission.
In an earlier work, it was suggested that the descendants of CCOs could be RPs, rotating radio transients (RRATs) or they may not be observable in any energy band \citep{Gencali2024}.
RRATs are INS systems that show repetitive short radio bursts with durations much shorter than their spin periods.
The physical mechanism behind these bursts is not well known \citep{Mc2006}. 
Penetration of the inner disc into the light cylinder could enhance the voltage across the magnetic poles \citep{Parfrey2016}.
It was suggested that this voltage enhancement by the field-disc interaction may trigger RRAT bursts in INS sources that would actually be dead pulsars if they had no discs \citep[see][for details]{Gencali2024}. 
From the preliminary results of our ongoing work, we estimate that CCOs could evolve into RPs, RRATs, or cool below current detection limits without showing any radio or X-ray emission (Ş. Demirok, Ü. Ertan, A. A. Gençali, in preparation). Using the results of these analyses, in the present work, we will focus on the long-term evolution of Calvera and CCOs with initial conditions similar to those of Calvera.
We will also discuss the possible descendants of these sources.
Unlike the three other CCOs with measured $P$ and $\Pdot$, Calvera lies well above the upper border of the pulsar death valley \citep{Chen1993}, while the source, like other CCOs, shows no regular radio pulsations \citep[see e.g.][]{DeLuca2017, Rigoselli2024}.
From our model results, we estimate that Calvera is spinning down by the disc torques while the mass accretion on to the NS is going on quenching the radio pulses at present. The source is likely to become an RP after the accretion is switched off. In Section~\ref{model}, we briefly describe the model. Our results are discussed in Section~\ref{results} and conclusions are summarized in Section~\ref{conc}.

\section{The Model}
\label{model}

Here, we briefly describe the long-term evolution model applied earlier to different INS populations \citep[see e.g.][]{Ertan2014, Benli2016} and the analytical torque model \citep{Ertan2017, Ertan2018, Ertan2021}. The mass accretion rate at the inner disc, $\Mdotin$, is determined by solving the disc diffusion equation for a geometrically thin and optically thick disc using the $\alpha$-prescription for the kinematic viscosity, $\nu = \alpha \cs h$, where $\alpha$ is the kinematic viscosity parameter, $\cs$ is the sound speed, and $h$ is the pressure scale height of the disc \citep{Shakura1973}.

The disc is heated by both viscous dissipation and X-ray irradiation. The X-ray irradiation flux can be written as $\Firr=1.2C \Lx/\upi r^2$, where $r$ is the radial distance from the star, $C$ is the irradiation efficiency parameter, $\Lx=\Lacc+\Lcool$ is the total X-ray luminosity \citep{Fukue1992}. In our previous work on the long-term evolution and evolutionary links of INS populations \citep{Gencali2024}, more consistent results were obtained with the maximal cooling curve of \citet{Page+2006,Page+2009} using the same analytical torque model.
For the cooling luminosity, $\Lcool$, of the NS, we use the same theoretical cooling curve in the present work. The accretion luminosity is related to the accretion rate on to the NS, $\Mdotstar$, through $\Lacc= GM\Mdotstar/r_{\ast}$, where $G$ is the gravitational constant, $M$ and $r_\ast$ are the mass and radius of the NS, for which we take $1.4~\Msun$ and $10^6$~cm respectively. The disc becomes viscously passive starting from the outermost regions when the effective temperature of the disc, $\Teff$, drops below a critical temperature, $\Tp$. The dynamic outer radius of the active disc, $\rout$, is defined as the radius at which $\Teff = \Tp$. During the long-term evolution, $\rout$ gradually decreases with decreasing $\Firr$. Eventually, the entire disc becomes viscously inactive and disc-field interaction terminates. 

The disc parameters $\alpha$, $\Tp$, and $C$ are expected to have similar values for different INS populations. In our earlier work, reasonable results were obtained with $\alpha=0.045$, $\Tp=50-150$~K, and $C=(1-7)\times10^{-4}$ \citep[see e.g.][]{Benli2016, Ertan+2017}. The initial conditions are the initial spin period, $P_0$, the initial disc mass, $\Md$, and $B_0$. The long-term evolution is most sensitive to $B_0$ \citep{Benli2016}, which is kept constant throughout the evolution, neglecting any magnetic field decay.
Different magnetic-field decay models \citep[e.g.][]{Geppert1996, Konar1999} estimate that the field decay is not significant until ages of $10^3-10^4$~yr for the $\Mdotstar$ ranges ($\sim 10^{13} - 10^{14}$~\gpers) estimated for CCOs. After the inactivation of the disc, a few $10^6$~yr, Ohmic decay of the dipole field is not significant either for the low $B_0$ values (a few $10^9 - 10^{10}$~G) estimated for CCOs in our model \citep{Pons2009}.

For the inner disc radius, $\rin$, and torque calculations, we use the analytical torque model developed by \citet{Ertan2017, Ertan2018, Ertan2021}.
This model is based on the basic principles of the model developed by \citet{Lovelace1995, Lovelace1999} and \citet{ Ustyugova2006}.
In this model, depending on the location of $\rin$ relative to the co-rotation radius, $\rco$, there are three different rotational phases, namely the strong propeller (SP) phase, the weak propeller (WP) phase, and the spin-up (SU) phase. 

In the SP phase, magnetic field lines and the inner disc interact in a narrow inner boundary region with radial width $\Delta r < r$. The field lines interacting with the matter in the boundary continually inflate, open up, throw the matter along the open field lines, and reconnect on dynamical timescale \citep{Lovelace1995, Ustyugova2006}. Outside the boundary region, the field lines are open and disconnected from the disc.
The maximum inner disc radius for which the system can sustain the strong propeller mechanism is estimated from
\begin{equation}        
        \Rinmax^{25/8}~|1 - \Rinmax^{-3/2}| ~ \simeq ~ 1.26  ~
\alpha_{-1}^{2/5} ~M_{1.4}^{-7/6} ~\Mdot_\mathrm{in,16}^{-7/20}~ \umu_{30} ~ P^{-13/12}
\label{eq:Rin}  
\end{equation}
\citep{Ertan2017, Ertan2018, Ertan2021}. Here, $\Rinmax = \rinmax / \rco$, $\alpha_{-1} = \alpha/0.1$, $M_{1.4} = M/1.4 \Msun$, $\Mdot_{{\rm in}, 16} = \Mdotin /10^{16}$~\gpers, and $\umu_{30} = \umu/10^{30}$~G~cm$^3$ is the magnetic dipole moment. The actual inner disc radius is estimated to be $\rin=\reta=\eta~\rinmax$ with $\eta \lesssim 1$, which is expected to be close to unity due to the strong radial dependence of the magnetic torques.

The total torque acting on the star can be written as
\begin{equation}        
        \Gamma = \Mdot_\ast\sqrt{GM\rin} -\frac{\mu^2}{\rin^3 }\Bigg(\frac{\Delta r}{\rin}\Bigg)-\frac{2\mu^2\Omega^3_\ast}{3c^3}
\label{eq:torque}  
\end{equation}
where the first term is the spin-up torque, $\Gammaacc$, produced by accretion on to the NS. The second term corresponds to the spin-down torque, $\GammaD$, arising from the interaction between the magnetic field lines and the inner disc inside a narrow interaction region with radial width $\Delta r < \rin$. The last term is the magnetic dipole torque, $\Gammadip$, where $\Omega_\ast$ is the angular velocity of the star and $c$ is the speed of light. 

At low $\Mdotin$ levels, the system is in the SP phase and the inflowing matter is thrown out from the narrow inner-disc boundary along the open field lines. 
A steady SP phase can be sustained only when $\rin > r_1 \simeq 1.26~\rco$, since the speed of the closed field lines co-rotating with the star exceeds the escape speed outside $r_1$.
In this phase, there is no mass accretion on to the NS ($\Mdotstar = 0$), $\Gamma = \GammaD + \Gammadip$, and $\Lx = \Lcool$. Sources with sufficient rotational power could emit ordinary radio pulsations during the SP phase.

At relatively high $\Mdotin$ levels, if $\rco<\rin<r_1$ instantaneously, the matter thrown out of the inner interaction region falls back to the disc at larger radii. This increases the surface densities at the inner disc, which pushes the inner disc inward until $\rin=\rco$. This initiates accretion on to the NS and the system makes a transition from the SP phase to the WP phase. The WP phase can persist for a wide range of $\Mdotin$. In this phase, NS spins down while mass flows from $\rco$ along the closed field lines on to the NS. In the WP phase, all the torques are active, $\Lx=\Lacc+\Lcool$, and ordinary radio pulsations are expected to be quenched due to ongoing mass accretion.

For $\Mdotin$ values above a critical value, the inner disc penetrates into $\rco$, which leads to WP/SU transition (torque reversal), which takes place when the viscous torques dominate the magnetic torques, at $\xi \rA \simeq r_\xi \simeq \rco$, where $\rA$ is the conventional \Alfven radius and $\xi \simeq 0.5 - 1$. As the $\Mdotstar$ levels of most INS populations remain below the critical level for the WP/SU transition during their long-term evolution \citep[see e.g.][]{Gencali2024}, the details of this phase will not be given here.

\section{Results and Discussion}
\label{results}

\begin{figure}
    \centering
    \includegraphics[width=\columnwidth]{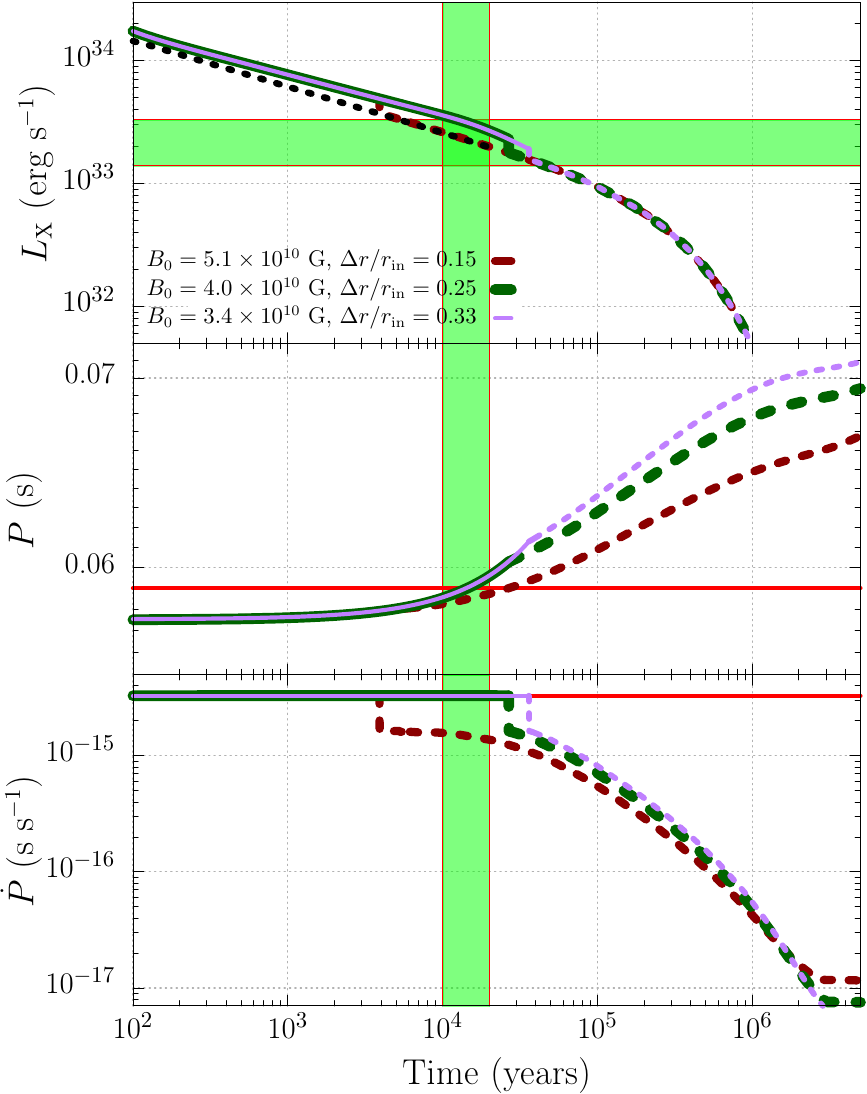}
    \caption{Illustrative model curves for the long-term evolution of \src~(Calvera). The results are obtained with $\alpha = 0.045$, $\Tp~=~50$~K, $C = 1 \times 10^{-4}$, $\eta=0.6$, $\xi = 0.7$, $P_0 \simeq 57.5$~ms, and $\Md \simeq 6.3 \times 10^{-7}~\Msun$. The thick green curves are obtained with $B_0 \simeq 4.0 \times 10^{10}$~G and $\DeltaR = 0.25$. The other two sets of curves are obtained with ($B_0$, $\DeltaR$) = ($5.1 \times 10^{10}$~G, $0.15$) and $(3.4 \times 10^{10}$~G, $0.33$), respectively, to illustrate the degeneracy between $B_0$ and $\DeltaR$ in $\GammaD$ calculation. The horizontal shaded area in the top panel shows the estimated $\Lx$ range corresponding to $d = 3-5$~kpc \citep{Mereghetti2021, Greco2025}. The solid lines in the middle and bottom panels show the measured $P$ and $\Pdot$ respectively. The vertical shaded area indicates the estimated SNR age range \citep{Greco2025}. The solid and dashed segments of the model curves correspond to the WP and the SP phases respectively. The dotted curve in the top panel shows the theoretical maximal cooling curve obtained by \citet{Page+2009}.
    }
    \label{fig:longterm_eta06}
\end{figure}

The model curve seen in Fig.~\ref{fig:longterm_eta06} illustrates a long-term evolution that can simultaneously reproduce the observed properties ($P$, $\Pdot$, and $\Lx$) of Calvera. Among the initial conditions ($B_0$, $P_0$, and $\Md$), our model results are most sensitive to $B_0$. 
We obtain reasonable results with $B_0 \simeq 4 \times 10^{10}$~G while the source is evolving in the WP phase at an age $\sim 10^4$~yr, and $\Lcool$ is the dominant source of $\Lx$. 
In this phase, due to mass accretion on to the star, ordinary radio pulses are not allowed in the model. This is in agreement with the lack of ordinary pulsed radio emission from Calvera, as well as from the other known CCOs.
Since the strength of $\GammaD$ acting on the source is relatively weak compared to those of other INS populations, we estimate that the $P_0$ ($\simeq 57.5$~ms) value is close to the observed $P$ ($\simeq 59.2$~ms). 
The $\Md$ values in the range $\sim (4 - 10) \times 10^{-7}~\Msun$ yield reasonable results consistent with the estimated SNR age. 
If the estimated $\Lx$ increases/decreases due to a distance correction, a similar model fit can be obtained by increasing/decreasing $\Md$, which does not change rotational evolution in the WP phase.

In the WP phase, the spin-down rate is mainly determined by the disc torque. Among the model parameters, $\DeltaR$ is degenerate with $B_0$ in the $\GammaD$ calculation. To illustrate this degeneracy, we added two additional model curves with different $B_0-\DeltaR$ pairs to Fig.~\ref{fig:longterm_eta06}, keeping the other parameters fixed. The $\rin$ value, which determines the current phase of the source, depends on $B_0$, but not on $\DeltaR$. Among these sources with the same $\GammaD$ in the WP phase, those with higher $B_0$ values make the WP/SP transition at higher $\Mdotin$ levels (at an earlier age; see Fig.~\ref{fig:longterm_eta06}).

\begin{figure*}
    \centering
    \includegraphics[width=\linewidth]{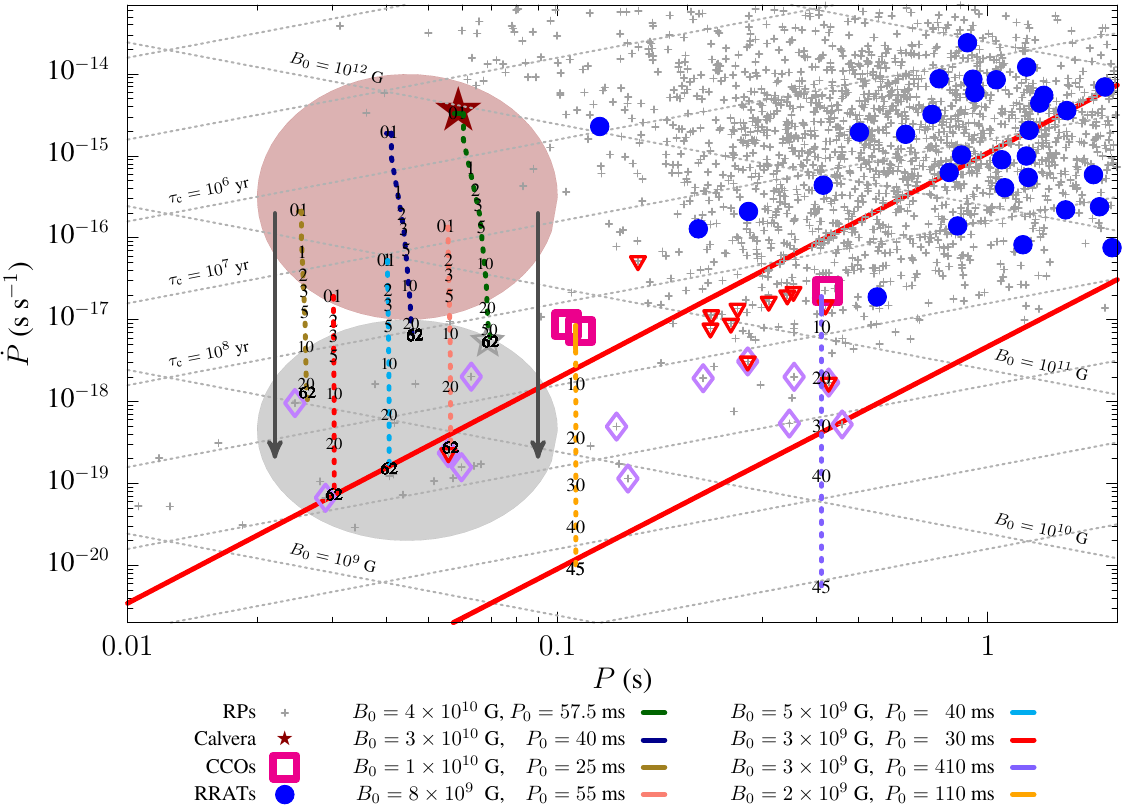}
    \caption{Long-term evolution of Calvera and illustrative sources with fallback disc in the $P-\Pdot$ diagram. For all model curves, $\alpha = 0.045$, $\Tp = 50$~K, $C = 1 \times 10^{-4}$, $\DeltaR=0.25$, $\eta=0.6$, $\xi=0.7$, and $\Md \simeq 6.3 \times 10^{-7}~\Msun$ (except for the two illustrative curves on the right for the three CCOs, for which $\eta = 0.7$ and $\Md \simeq 1.6 \times 10^{-6}~\Msun$; from \citet{Gencali2024}). The $B_0$ and $P_0$ values are given below the figure. Solid and dashed curves correspond to the WP and SP phases respectively. Diagonal solid lines are the upper and lower boundaries of the pulsar death valley \citep{Chen1993}. The constant-$\tau_{\rm c}$ and $B_0$ lines calculated from the magnetic dipole formula are also given for comparison. The numbers on the model curves show the ages of the sources in units of $10^5$~yr. The data for RPs and CCOs are taken from the ATNF Pulsar Catalogue \citep[version 2.7.0;][]{Manchester2005}, while those for RRATs are taken from \textsc{RRATalog}\protect\footnote{\url{https://rratalog.github.io/rratalog}} \citep{Agarwal2026}. Empty diamonds and triangles denote the RPs investigated in X-rays by \citet{Gotthelf2013DRP} and \citet{Luo2015} respectively. Filled and empty star show the current and final locations of Calvera in the model. The shaded areas denote the approximate regions occupied by Calvera-like CCOs at birth and at the end of their evolution. Two arrows indicate the evolutionary direction of the sources (see the text for details).}
    \label{fig:PPdot}
\end{figure*}

For the model curves seen in Fig.~\ref{fig:longterm_eta06}, the source will remain in the WP phase for $\sim 2 \times 10^4$~yr. Subsequently, it will enter the SP phase (dashed curve) and start emitting ordinary radio pulses since the source will have sufficient rotational power for this emission and the accretion is not allowed in this phase.
In the SP phase, the $\GammaD$ and $\Pdot$ decrease with decreasing $\Mdotin$, and eventually weak $\Gammadip$ dominates $\GammaD$.
After the inactivation of the entire disc, the source will slow down by $\Gammadip$ only as an RP in the same region of the $P-\Pdot$ diagram for a very long time due to weak $\Gammadip$ (empty star in Fig.~\ref{fig:PPdot}).

In the earlier work, the long-term evolution and evolutionary links of the INS populations, including CCOs, have been investigated in detail \citep{Gencali2024}. The three CCOs with known $P$ and $\Pdot$ values are estimated to be evolving at points below or inside the death valley (delineated by the red diagonal solid lines in Fig.~\ref{fig:PPdot}) and very close to the lower border when the WP phase terminates. Therefore, the other CCOs within the $P$ and $\Pdot$ ranges of these three sources are not likely to become RPs even after the termination of the accretion phase.

The location of Calvera on the $P - \Pdot$ diagram (filled star in Fig.~\ref{fig:PPdot}) is rather different from those of the other three CCOs. Indeed, it could not be identified as a CCO until the discovery of the SNR around the source. This could indicate that there could be CCOs starting their evolutions with the initial conditions similar to that of Calvera. Pursuing this idea, we have investigated the long-term evolution of the sources starting their evolution within the red region seen in Fig.~\ref{fig:PPdot}. We have chosen some illustrative initial conditions in the red region. 
The red region in Fig.~\ref{fig:PPdot} corresponds to a parameter space with a $B_0$ range of $3 \times 10^9 - 1 \times 10^{11}$~G and $P_0$ range of $20-100$~ms. The long-term rotational evolution of the sources is not very sensitive to $\Md$ \citep{Benli2016}. We have obtained all these model curves with $\Md \simeq 6.3 \times 10^{-7}~M_\odot$. The evolution of these illustrative model sources with active disc terminates in the grey region as RPs. If the end point of the evolution is inside the death valley, the source could become an RP depending on the location of its death point inside the valley. It is seen in Fig.~\ref{fig:PPdot} that most of the sources initially in the red region find themselves either above the upper border of the death valley or inside and close to the upper border of the valley. This means that a large fraction of the CCOs born in the red region are likely to become RPs after termination of the accretion in this model.
After the inactivation of their discs, they are estimated to remain observable as RPs without a significant change in their $P$ and $\Pdot$ for a very long time ($>10^8$~yr). Indeed, there is a cluster of RPs in the grey region. Some of these sources were proposed earlier to be DRPs, which are NSs mildly recycled in HMXBs. Our results imply that some of these RPs could be descendants of CCOs, initially similar to Calvera. In our picture, Calvera is a very young source, while the descendant RPs (in the grey region) are likely to be very old sources. This means that their birth rate (Calvera-like CCOs) should be low considering that they should remain in the grey region as an RP for a long timescale. This is likely to be the reason for the lack of other CCOs in the red region around Calvera. Some of the RPs in the grey region could be the old ($\tau > 10^8$~yr) descendants of CCOs. 
In this region, which includes $\sim 15$ old RPs, it is very unlikely to detect immediate descendants of CCOs in X-rays considering that their cooling luminosity could be detected at ages $\sol 10^6$~yr. Among these RPs, five sources were investigated in X-rays, and no X-ray emission was detected \citep{Gotthelf2013DRP, Luo2015}.

For an order of magnitude estimate, let us assume that $10$ out of $15$ RPs in the grey region are descendants of Calvera-like CCOs with lifetimes $\sim 10^8$~yr. This gives a birth rate of $1 \times 10^{-7}$~yr$^{-1}$. Taking the lifetime observable through cooling luminosity $\sim 10^6$~yr and the RP lifetime $\sim 10^8$~yr, the probability of observing a Calvera-like source in this sample is $\sim 1/10$. This is an interesting coincidence if there are indeed $\sim 10$ Calvera-like sources among which only Calvera is detectable in X-rays. Note that the minimum birth rate of all CCOs estimated from SNR expansion time and the total number of CCOs is $\sim 0.04$~century$^{-1}$ \citep{Kaspi2010}, much higher than we estimated for the Calvera-like CCOs.

\section{Conclusions}
\label{conc}
We have investigated the long-term evolution and possible descendants of Calvera-like CCOs with the model applied earlier to other INS populations.
We have shown that a NS evolving with a fallback disc and a dipole field strength $B_0 \simeq 4 \times 10^{10}$~G can achieve the observed properties ($P$, $\Pdot$, and $\Lx$) of Calvera consistently with its estimated SNR age ($\sim 10^4$~yr). In the model, Calvera is evolving in the WP phase at present, and the accretion on to the star in this phase can account for the lack of radio pulses from the source. 
Our simulations indicate that the source will become an RP after the termination of the WP phase at a time $\tau \simeq 2 \times 10^4$~yr. 
The source will remain as an RP for a very long time ($\gtrsim 10^8$~yr) due to weak $\Gammadip$ that will act on the source after the inactivation of the disc at an age $\tau\sim {\rm a~few~} \times 10^6$~yr.
Our simulations show that Calvera and similar CCOs evolve vertically downward in the $P-\Pdot$ diagram and remain above or close to the upper border of the death valley emerging as RPs. These descendants of CCOs are estimated to have a very long RP life-time.
Indeed, there is a cluster of RPs ($\sim 15$ RPs) in the region into which Calvera-like model sources evolve in our model.
Inside this old population we do not expect to find young immediate descendants of CCOs that could be detected with their cooling luminosity in X-rays.
Among these, five sources were observed, but could not be detected in X-rays \citep{Gotthelf2013DRP, Luo2015}.

\section*{Acknowledgements}

We thank the referee for their useful comments, which have considerably improved our manuscript. We acknowledge research support from Sabanc{\i} University, and from T\"{U}B\.{I}TAK (The Scientific and Technological Research Council of Turkey) through grant 125F197. We thank M. Ali Alpar for useful comments on the manuscript.

\section*{Data Availability}
No new data were analysed in support of this paper.



\bibliographystyle{mnras}
\bibliography{example} 

@ARTICLE{Alpar2001,
   author = {{Alpar}, M.~A.},
    title = "{On Young Neutron Stars as Propellers and Accretors with Conventional Magnetic Fields}",
  journal = {\apj},
   eprint = {astro-ph/0005211},
     year = 2001,
    month = jun,
   volume = 554,
    pages = {1245-1254},
      doi = {10.1086/321393},
   adsurl = {http://adsabs.harvard.edu/abs/2001ApJ...554.1245A}
}

@ARTICLE{Agarwal2026,
       author = {{Agarwal}, Devansh and {Lewis}, Evan F. and {Lorimer}, Duncan R. and {McLaughlin}, Maura A. and {Cui}, Bingyi and {Turner}, Anna and {McMann}, Natasha},
        title = "{The RRATALOG: a Galactic census of rotating radio transients}",
      journal = {\mnras},
         year = 2026,
        month = jun,
       volume = {549},
       number = {1},
          eid = {stag787},
        pages = {stag787},
          doi = {10.1093/mnras/stag787},
archivePrefix = {arXiv},
       eprint = {2604.01203},
 primaryClass = {astro-ph.HE},
       adsurl = {https://ui.adsabs.harvard.edu/abs/2026MNRAS.549ag787A}
}

@article{Arias2022,
   author = {M. Arias and A. Botteon and C. G. Bassa and S. Van Der Jagt and R. J. Van Weeren and S. P. Oasullivan and Q. Bosschaart and R. S. Dullaart and M. J. Hardcastle and J. W.T. Hessels and T. Shimwell and M. M. Slob and J. A. Sturm and C. Tasse and N. C.M.A. Theijssen and J. Vink},
   doi = {10.1051/0004-6361/202244369},
   issn = {0004-6361},
   journal = {\aap},
   month = {11},
   pages = {A71},
   publisher = {EDP Sciences},
   title = {Possible discovery of Calvera’s supernova remnant},
   volume = {667},
   url = {https://www.aanda.org/articles/aa/full_html/2022/11/aa44369-22/aa44369-22.html https://www.aanda.org/articles/aa/abs/2022/11/aa44369-22/aa44369-22.html},
   year = {2022}
}

@article{Araya2023,
   author = {M Araya},
   doi = {10.1093/mnras/stac3337},
   journal = {MNRAS},
   pages = {4132-4137},
   title = {Fermi-LAT detection of G118.4 + 37.0: a superno v a remnant in the Galactic halo seen around the Calvera pulsar},
   volume = {518},
   url = {https://doi.org/10.1093/mnras/stac3337},
   year = {2023}
}

@ARTICLE{Chevalier1989,
       author = {{Chevalier}, Roger A.},
        title = "{Neutron Star Accretion in a Supernova}",
      journal = {\apj},
         year = 1989,
        month = nov,
       volume = {346},
        pages = {847},
          doi = {10.1086/168066},
       adsurl = {https://ui.adsabs.harvard.edu/abs/1989ApJ...346..847C}
}

@ARTICLE{Colgate1971,
       author = {{Colgate}, Stirling A.},
        title = "{Neutron-Star Formation, Thermonuclear Supernovae, and Heavy-Element Reimplosion}",
      journal = {\apj},
         year = 1971,
        month = jan,
       volume = {163},
        pages = {221},
          doi = {10.1086/150760},
       adsurl = {https://ui.adsabs.harvard.edu/abs/1971ApJ...163..221C}
}

@ARTICLE{Chatterjee2000,
   author = {{Chatterjee}, P. and {Hernquist}, L. and {Narayan}, R.},
    title = "{An Accretion Model for Anomalous X-Ray Pulsars}",
  journal = {\apj},
   eprint = {astro-ph/9912137},
     year = 2000,
    month = may,
   volume = 534,
    pages = {373-379},
      doi = {10.1086/308748},
   adsurl = {http://adsabs.harvard.edu/abs/2000ApJ...534..373C}
}

@ARTICLE{Chen1993,
       author = {{Chen}, Kaiyou and {Ruderman}, Malvin},
        title = "{Pulsar Death Lines and Death Valley}",
      journal = {\apj},
         year = 1993,
        month = jan,
       volume = {402},
        pages = {264},
          doi = {10.1086/172129},
       adsurl = {https://ui.adsabs.harvard.edu/abs/1993ApJ...402..264C}
}

@ARTICLE{Geppert1999,
       author = {{Geppert}, Ulrich and {Page}, Dany and {Zannias}, Thomas},
        title = "{Submergence and re-diffusion of the neutron star magnetic field after the supernova}",
      journal = {\aap},
         year = 1999,
        month = may,
       volume = {345},
        pages = {847-854},
       adsurl = {https://ui.adsabs.harvard.edu/abs/1999A&A...345..847G}
}

@article{Daki2025,
   author = {V. Dakić and S. B. Popov and R. Turolla},
   doi = {10.1051/0004-6361/202555186},
   issn = {0004-6361},
   journal = {\aap},
   month = {9},
   pages = {A21},
   publisher = {EDP Sciences},
   title = {Supernova explosions of runaway stars and young neutron stars above the Galactic plane},
   volume = {701},
   url = {https://www.aanda.org/articles/aa/full_html/2025/09/aa55186-25/aa55186-25.html https://www.aanda.org/articles/aa/abs/2025/09/aa55186-25/aa55186-25.html},
   year = {2025}
}

@article{DeLuca2017,
doi = {10.1088/1742-6596/932/1/012006},
url = {https://doi.org/10.1088/1742-6596/932/1/012006},
year = {2017},
month = {dec},
publisher = {IOP Publishing},
volume = {932},
number = {1},
pages = {012006},
author = {De Luca, A},
title = {Central compact objects in supernova remnants},
journal = {Journal of Physics: Conference Series}
}

@ARTICLE{Benli2016,
   author = {{Benli}, O. and {Ertan}, {\"U}.},
    title = "{Long-term evolution of anomalous X-ray pulsars and soft gamma repeaters}",
  journal = {\mnras},
archivePrefix = "arXiv",
   eprint = {1601.07846},
 primaryClass = "astro-ph.HE",
     year = 2016,
    month = apr,
   volume = 457,
    pages = {4114-4122},
      doi = {10.1093/mnras/stw235},
   adsurl = {http://adsabs.harvard.edu/abs/2016MNRAS.457.4114B}
}

@ARTICLE{Benli2017,
   author = {{Benli}, O. and {Ertan}, {\"U}.},
    title = "{On the evolution of high-B radio pulsars with measured braking indices}",
  journal = {\mnras},
archivePrefix = "arXiv",
   eprint = {1707.02100},
 primaryClass = "astro-ph.HE",
     year = 2017,
    month = nov,
   volume = 471,
    pages = {2553-2557},
      doi = {10.1093/mnras/stx1735},
   adsurl = {http://adsabs.harvard.edu/abs/2017MNRAS.471.2553B}
}

@ARTICLE{Benli2018,
       author = {{Benli}, Onur and {Ertan}, {\"U}nal},
        title = "{Rotational and X-ray luminosity evolution of high-B radio pulsars}",
      journal = {\na},
         year =  {2018},
        month = "May",
       volume = {61},
        pages = {78-83},
          doi = {10.1016/j.newast.2017.12.005},
archivePrefix = {arXiv},
       eprint = {1712.06505},
 primaryClass = {astro-ph.HE},
       adsurl = {https://ui.adsabs.harvard.edu/abs/2018NewA...61...78B}
}

@ARTICLE{BenliCCO2018,
    author = {{Benli}, O. and {Ertan}, {\"U}.},
    title = "{Central Compact Objects: some of them could be spinning up?}",
    journal = {\mnras},
    volume = {478},
    number = {4},
    pages = {4890-4893},
    year = {2018},
    month = {05},
    issn = {0035-8711},
    doi = {10.1093/mnras/sty1399},
    url = {https://doi.org/10.1093/mnras/sty1399},
    eprint = {https://academic.oup.com/mnras/article-pdf/478/4/4890/25100621/sty1399.pdf},
}

@ARTICLE{Ertan2006,
   author = {{Ertan}, {\"U}. and {{\c C}al{\i}{\c s}kan}, {\c S}.},
    title = "{Optical and Infrared Emission from the Anomalous X-Ray Pulsars and Soft Gamma-Ray Repeaters}",
  journal = {ApJ},
   eprint = {astro-ph/0608288},
     year = 2006,
    month = oct,
   volume = 649,
    pages = {L87-L90},
      doi = {10.1086/508347},
   adsurl = {http://adsabs.harvard.edu/abs/2006ApJ...649L..87E}
}

@article{Ertan2007,
	author = {{\"U}. Ertan and M. H. Erkut and K. Y. Ek{\c{s}}i and M. A. Alpar},
    title = "{The Anomalous X-Ray Pulsar 4U 0142+61: A Neutron Star with a Gaseous Fallback Disk}",
	journal = {\apj},
	year = 2007,
	month = mar,
	volume = {657},
	number = {1},
	pages = {441-447},
      doi = {10.1086/510303},
archivePrefix = {arXiv},
       eprint = {astro-ph/0612587},
 primaryClass = {astro-ph},
       adsurl = {https://ui.adsabs.harvard.edu/abs/2007ApJ...657..441E}
}

@ARTICLE{Ertan2009,
   author = {{Ertan}, {\"U}. and {Ek{\c s}i}, K.~Y. and {Erkut}, M.~H. and 
	{Alpar}, M.~A.},
    title = "{On the Evolution of Anomalous X-ray Pulsars and Soft Gamma-ray Repeaters with Fall Back Disks}",
  journal = {\apj},
archivePrefix = "arXiv",
   eprint = {0907.3222},
 primaryClass = "astro-ph.HE",
     year = 2009,
    month = sep,
   volume = 702,
    pages = {1309-1320},
      doi = {10.1088/0004-637X/702/2/1309},
   adsurl = {http://adsabs.harvard.edu/abs/2009ApJ...702.1309E}
}

@ARTICLE{Ertan2014,
   author = {{Ertan}, {\"U}. and {{\c C}al{\i}{\c s}kan}, {\c S}. and {Benli}, O. and 
	{Alpar}, M.~A.},
    title = "{Long-term evolution of dim isolated neutron stars}",
  journal = {\mnras},
archivePrefix = "arXiv",
   eprint = {1408.0650},
 primaryClass = "astro-ph.HE",
     year = 2014,
    month = oct,
   volume = 444,
    pages = {1559-1565},
      doi = {10.1093/mnras/stu1523},
   adsurl = {http://adsabs.harvard.edu/abs/2014MNRAS.444.1559E}
}

@ARTICLE{Ertan2017,
   author = {{Ertan}, {\"U}.},
    title = "{The inner disc radius in the propeller phase and accretion-propeller transition of neutron stars}",
  journal = {\mnras},
     year = 2017,
    month = apr,
   volume = 466,
    pages = {175-180},
      doi = {10.1093/mnras/stw3131},
   adsurl = {http://adsabs.harvard.edu/abs/2017MNRAS.466..175E}
}

@ARTICLE{Ertan+2017,
       author = {{Ertan}, {\"U}. and {{\c{C}}al{\i}{\c{s}}kan}, {\c{S}}. and {Alpar}, M.~A.},
        title = "{Optical excess of dim isolated neutron stars}",
      journal = {\mnras},
         year = 2017,
        month = may,
       volume = {470},
       number = {1},
        pages = {1253-1258},
          doi = {10.1093/mnras/stx1310},
archivePrefix = {arXiv},
       eprint = {1705.09547},
 primaryClass = {astro-ph.HE},
       adsurl = {https://ui.adsabs.harvard.edu/abs/2017MNRAS.470.1253E}
}

@ARTICLE{Ertan2018,
   author = {{Ertan}, {\"U}.},
    title = "{Accretion and propeller torque in the spin-down phase of neutron stars: The case of transitional millisecond pulsar PSR J1023+0038}",
  journal = {\mnras},
archivePrefix = "arXiv",
   eprint = {1805.10091},
 primaryClass = "astro-ph.HE",
     year = 2018,
    month = sep,
   volume = 479,
    pages = {L12-L16},
      doi = {10.1093/mnrasl/sly089},
   adsurl = {http://adsabs.harvard.edu/abs/2018MNRAS.479L..12E}
}

@ARTICLE{Ertan2021,
       author = {{Ertan}, {\"U}nal},
        title = "{On the torque reversals of accreting neutron stars}",
      journal = {\mnras},
         year = 2021,
        month = jan,
       volume = {500},
       number = {3},
        pages = {2928-2936},
          doi = {10.1093/mnras/staa3378},
archivePrefix = {arXiv},
       eprint = {2010.15035},
 primaryClass = {astro-ph.HE},
       adsurl = {https://ui.adsabs.harvard.edu/abs/2021MNRAS.500.2928E}
}

@ARTICLE{Fukue1992,
   author = {{Fukue}, J.},
    title = "{Self-irradiated accretion disks}",
  journal = {\pasj},
     year = 1992,
    month = dec,
   volume = 44,
    pages = {663-667},
   adsurl = {http://adsabs.harvard.edu/abs/1992PASJ...44..663F}
}

@ARTICLE{Gencali2018,
       author = {{Gen{\c{c}}ali}, A.~A. and {Ertan}, {\"U}.},
        title = "{Long-term evolution of RRAT J1819-1458}",
      journal = {\mnras},
         year =  2018,
        month = "Nov",
       volume = {481},
       number = {1},
        pages = {244-249},
          doi = {10.1093/mnras/sty2287},
archivePrefix = {arXiv},
       eprint = {1808.06053},
 primaryClass = {astro-ph.HE},
       adsurl = {https://ui.adsabs.harvard.edu/abs/2018MNRAS.481..244G}
}

@ARTICLE{Gotthelf2005,
       author = {{Gotthelf}, E.~V. and {Halpern}, J.~P. and {Seward}, F.~D.},
        title = "{Discovery of a 105 ms X-Ray Pulsar in Kesteven 79: On the Nature of Compact Central Objects in Supernova Remnants}",
      journal = {\apj},
         year = 2005,
        month = jul,
       volume = {627},
       number = {1},
        pages = {390-396},
          doi = {10.1086/430300},
archivePrefix = {arXiv},
       eprint = {astro-ph/0503424},
 primaryClass = {astro-ph},
       adsurl = {https://ui.adsabs.harvard.edu/abs/2005ApJ...627..390G}
}

@article{Ho2011,
    author = {Ho, Wynn C. G.},
    title = {Evolution of a buried magnetic field in the central compact object neutron stars},
    journal = {\mnras},
    volume = {414},
    number = {3},
    pages = {2567-2575},
    year = {2011},
    month = {06},
    issn = {0035-8711},
    doi = {10.1111/j.1365-2966.2011.18576.x},
    url = {https://doi.org/10.1111/j.1365-2966.2011.18576.x},
    eprint = {https://academic.oup.com/mnras/article-pdf/414/3/2567/3535294/mnras0414-2567.pdf},
}

@article{Vigano2013hiddenB,
    author = {Viganò, D. and Pons, J. A.},
    title = {Central compact objects and the hidden magnetic field scenario},
    journal = {\mnras},
    volume = {425},
    number = {4},
    pages = {2487-2492},
    year = {2012},
    month = {10},
    issn = {0035-8711},
    doi = {10.1111/j.1365-2966.2012.21679.x},
    url = {https://doi.org/10.1111/j.1365-2966.2012.21679.x},
    eprint = {https://academic.oup.com/mnras/article-pdf/425/4/2487/4880432/425-4-2487.pdf},
}

@ARTICLE{Geppert1996,
       author = {{Geppert}, U. and {Urpin}, V. and {Konenkov}, D.},
        title = "{Wind accretion and magnetorotational evolution of neutron stars in binaries.}",
      journal = {\aap},
         year = 1996,
        month = mar,
       volume = {307},
        pages = {807-812},
       adsurl = {https://ui.adsabs.harvard.edu/abs/1996A&A...307..807G}
}

@article{Gotthelf2013DRP,
doi = {10.1088/0004-637X/773/2/141},
url = {https://dx.doi.org/10.1088/0004-637X/773/2/141},
year = {2013},
month = {aug},
publisher = {The American Astronomical Society},
volume = {773},
number = {2},
pages = {141},
author = {Gotthelf, E. V. and Halpern, J. P. and Allen, B. and Knispel, B.},
title = {X-RAY OBSERVATIONS OF DISRUPTED RECYCLED PULSARS: NO REFUGE FOR ORPHANED CENTRAL COMPACT OBJECTS},
journal = {\apj}
}

@article{Gotthelf2013CCOs,
doi = {10.1088/0004-637X/765/1/58},
url = {https://dx.doi.org/10.1088/0004-637X/765/1/58},
year = {2013},
month = {feb},
publisher = {The American Astronomical Society},
volume = {765},
number = {1},
pages = {58},
author = {Gotthelf, E. V. and Halpern, J. P. and Alford, J.},
title = {THE SPIN-DOWN OF PSR J0821–4300 AND PSR J1210–5226: CONFIRMATION OF CENTRAL COMPACT OBJECTS AS ANTI-MAGNETARS},
journal = {\apj}
}

@article{Luo2015,
doi = {10.1088/0004-637X/808/2/130},
url = {https://dx.doi.org/10.1088/0004-637X/808/2/130},
year = {2015},
month = {jul},
publisher = {The American Astronomical Society},
volume = {808},
number = {2},
pages = {130},
author = {Luo, J. and Ng, C.-Y. and Ho, W. C. G. and Bogdanov, S. and Kaspi, V. M. and He, C.},
title = {HUNTING FOR ORPHANED CENTRAL COMPACT OBJECTS AMONG RADIO PULSARS},
journal = {\apj}
}

@article{Lorimer2004,
    author = {Lorimer, D. R. and McLaughlin, M. A. and Arzoumanian, Z. and Xilouris, K. M. and Cordes, J. M. and Lommen, A. N. and Fruchter, A. S. and Chandler, A. M. and Backer, D. C.},
    title = {PSR J0609+2130: a disrupted binary pulsar?},
    journal = {\mnras},
    volume = {347},
    number = {2},
    pages = {L21-L25},
    year = {2004},
    month = {01},
    issn = {0035-8711},
    doi = {10.1111/j.1365-2966.2004.07407.x},
    url = {https://doi.org/10.1111/j.1365-2966.2004.07407.x},
    eprint = {https://academic.oup.com/mnras/article-pdf/347/2/L21/4891168/347-2-L21.pdf},
}

@article{Parfrey2016,
   title={TORQUE ENHANCEMENT, SPIN EQUILIBRIUM, AND JET POWER FROM DISK-INDUCED OPENING OF PULSAR MAGNETIC FIELDS},
   volume={822},
   ISSN={1538-4357},
   url={http://dx.doi.org/10.3847/0004-637X/822/1/33},
   DOI={10.3847/0004-637x/822/1/33},
   number={1},
   journal={\apj},
   publisher={American Astronomical Society},
   author={Parfrey, Kyle and Spitkovsky, Anatoly and Beloborodov, Andrei M.},
   year={2016},
   month=may, pages={33} }

@ARTICLE{Gencali2021,
       author = {{Gen{\c{c}}ali}, A.~A. and {Ertan}, {\"U}.},
        title = "{On the long-term evolution of rotating radio transients}",
      journal = {\mnras},
         year = 2021,
        month = jan,
       volume = {500},
       number = {3},
        pages = {3281-3289},
          doi = {10.1093/mnras/staa3371},
archivePrefix = {arXiv},
       eprint = {2010.14191},
 primaryClass = {astro-ph.HE},
       adsurl = {https://ui.adsabs.harvard.edu/abs/2021MNRAS.500.3281G}
}

@ARTICLE{Gencali2022,
       author = {{Gen{\c{c}}ali}, A.~A. and {Ertan}, {\"U}. and {Alpar}, M.~A.},
        title = "{Evolution of the long-period pulsar GLEAM-X J162759.5-523504.3}",
      journal = {\mnras},
         year = {2022},
        month = jun,
       volume = {513},
       number = {1},
        pages = {L68-L71},
          doi = {10.1093/mnrasl/slac034},
archivePrefix = {arXiv},
       eprint = {2202.06852},
 primaryClass = {astro-ph.HE},
       adsurl = {https://ui.adsabs.harvard.edu/abs/2022MNRAS.513L..68G}
}

@ARTICLE{Kaspi2010,
       author = {{Kaspi}, Victoria M.},
        title = "{Grand unification of neutron stars}",
      journal = {Proceedings of the National Academy of Science},
         year = 2010,
        month = apr,
       volume = {107},
       number = {16},
        pages = {7147-7152},
          doi = {10.1073/pnas.1000812107},
archivePrefix = {arXiv},
       eprint = {1005.0876},
 primaryClass = {astro-ph.HE},
       adsurl = {https://ui.adsabs.harvard.edu/abs/2010PNAS..107.7147K}
}

@ARTICLE{Gencali2023,
       author = {{Gen{\c{c}}ali}, A.~A. and {Ertan}, {\"U}. and {Alpar}, M.~A.},
        title = "{Evolution of the long-period pulsar PSR J0901-4046}",
      journal = {\mnras},
         year = {2023},
        month = mar,
       volume = {520},
       number = {1},
        pages = {L11-L15},
          doi = {10.1093/mnrasl/slac164},
archivePrefix = {arXiv},
       eprint = {2212.10501},
 primaryClass = {astro-ph.HE},
       adsurl = {https://ui.adsabs.harvard.edu/abs/2023MNRAS.520L..11G}
}

@BOOK{Tauris2023,
       author = {{Tauris}, Thomas M. and {van den Heuvel}, Edward P.~J.},
        title = "{Physics of Binary Star Evolution. From Stars to X-ray Binaries and Gravitational Wave Sources}",
         year = 2023,
          doi = {10.48550/arXiv.2305.09388},
       adsurl = {https://ui.adsabs.harvard.edu/abs/2023pbse.book.....T}
}

@article{Geppert2013,
   title={Radio pulsar activity and the crustal Hall drift},
   volume={435},
   ISSN={0035-8711},
   url={http://dx.doi.org/10.1093/mnras/stt1527},
   DOI={10.1093/mnras/stt1527},
   number={4},
   journal={\mnras},
   publisher={Oxford University Press (OUP)},
   author={Geppert, U. and Gil, J. and Melikidze, G.},
   year={2013},
   month=sep, pages={3262–3271} }

@ARTICLE{Gencali2024,
       author = {{Gen{\c{c}}ali}, A.~A. and {Ertan}, {\"U}.},
        title = "{Long-term evolutionary links between the isolated neutron star populations}",
      journal = {\mnras},
         year = 2024,
        month = oct,
       volume = {534},
       number = {2},
        pages = {1481-1489},
          doi = {10.1093/mnras/stae2177},
archivePrefix = {arXiv},
       eprint = {2409.11595},
 primaryClass = {astro-ph.HE},
       adsurl = {https://ui.adsabs.harvard.edu/abs/2024MNRAS.534.1481G}
}

@article{Greco2025,
   author = {Emanuele Greco and Michela Rigoselli and Sandro Mereghetti and Fabrizio Bocchino and Marco Miceli and Vincenzo Sapienza and Salvatore Orlando},
   doi = {10.1051/0004-6361/202555640},
   isbn = {0940970201},
   issn = {0004-6361},
   journal = {\aap},
   month = {9},
   pages = {A43},
   publisher = {EDP Sciences},
   title = {Multi-wavelength study of the high Galactic latitude supernova remnant candidate G118.4+37.0 associated with the Calvera pulsar},
   volume = {701},
   url = {https://www.aanda.org/articles/aa/full_html/2025/09/aa55640-25/aa55640-25.html https://www.aanda.org/articles/aa/abs/2025/09/aa55640-25/aa55640-25.html},
   year = {2025}
}

@ARTICLE{Haberl2007,
       author = {{Haberl}, Frank},
        title = "{The magnificent seven: magnetic fields and surface temperature distributions}",
      journal = {\apss},
         year = 2007,
        month = apr,
       volume = {308},
       number = {1-4},
        pages = {181-190},
          doi = {10.1007/s10509-007-9342-x},
archivePrefix = {arXiv},
       eprint = {astro-ph/0609066},
 primaryClass = {astro-ph},
       adsurl = {https://ui.adsabs.harvard.edu/abs/2007Ap&SS.308..181H}
}

@ARTICLE{Halpern2010,
       author = {{Halpern}, J.~P. and {Gotthelf}, E.~V.},
        title = "{Spin-Down Measurement of PSR J1852+0040 in Kesteven 79: Central Compact Objects as Anti-Magnetars}",
      journal = {\apj},
         year = 2010,
        month = jan,
       volume = {709},
       number = {1},
        pages = {436-446},
          doi = {10.1088/0004-637X/709/1/436},
archivePrefix = {arXiv},
       eprint = {0911.0093},
 primaryClass = {astro-ph.HE},
       adsurl = {https://ui.adsabs.harvard.edu/abs/2010ApJ...709..436H}
}

@article{Halpern2013,
   author = {J. P. Halpern and S. Bogdanov and E. V. Gotthelf},
   doi = {10.1088/0004-637X/778/2/120},
   issn = {0004-637X},
   issue = {2},
   journal = {\apj},
   month = {11},
   pages = {120},
   publisher = {IOP Publishing},
   title = {X-RAY MEASUREMENT OF THE SPIN-DOWN OF CALVERA: A RADIO- AND GAMMA-RAY-QUIET PULSAR},
   volume = {778},
   url = {https://iopscience.iop.org/article/10.1088/0004-637X/778/2/120 https://iopscience.iop.org/article/10.1088/0004-637X/778/2/120/meta},
   year = {2013}
}

@ARTICLE{Konar1999,
       author = {{Konar}, Sushan and {Bhattacharya}, Dipankar},
        title = "{Magnetic field evolution of accreting neutron stars - II}",
      journal = {\mnras},
         year = 1999,
        month = mar,
       volume = {303},
       number = {3},
        pages = {588-594},
          doi = {10.1046/j.1365-8711.1999.02287.x},
archivePrefix = {arXiv},
       eprint = {astro-ph/9808119},
 primaryClass = {astro-ph},
       adsurl = {https://ui.adsabs.harvard.edu/abs/1999MNRAS.303..588K}
}

@ARTICLE{Kylafis2014,
   author = {{Kylafis}, N.~D. and {Tr{\"u}mper}, J.~E. and {Ertan}, {\"U}.
	},
    title = "{Spectral formation in a radiative shock: application to anomalous X-ray pulsars and soft gamma-ray repeaters}",
  journal = {\aap},
archivePrefix = "arXiv",
   eprint = {1312.7282},
 primaryClass = "astro-ph.HE",
     year = 2014,
    month = feb,
   volume = 562,
      eid = {A62},
    pages = {A62},
      doi = {10.1051/0004-6361/201322303},
   adsurl = {http://adsabs.harvard.edu/abs/2014A%26A...562A..62K}
}

@ARTICLE{Lorimer2008,
       author = {{Lorimer}, Duncan R.},
        title = "{Binary and Millisecond Pulsars}",
      journal = {Living Reviews in Relativity},
         year = 2008,
        month = dec,
       volume = {11},
       number = {1},
          eid = {8},
        pages = {8},
          doi = {10.12942/lrr-2008-8},
archivePrefix = {arXiv},
       eprint = {0811.0762},
 primaryClass = {astro-ph},
       adsurl = {https://ui.adsabs.harvard.edu/abs/2008LRR....11....8L}
}

@article{Lovelace1995,
    title = {{Spin-up/spin-down of magnetized stars with accretion discs and outflows}},
    year = {1995},
    journal = {\mnras},
    author = {Lovelace, R. V. E. and Romanova, M. M. and Bisnovatyi-Kogan, G. S.},
    number = {2},
    month = {7},
    pages = {244--254},
    volume = {275},
    url = {https://academic.oup.com/mnras/article-lookup/doi/10.1093/mnras/275.2.244},
    doi = {10.1093/mnras/275.2.244},
    issn = {0035-8711}
}

@article{Lovelace1999,
    title = {{Magnetic Propeller Outflows}},
    year = {1999},
    journal = {\apj},
    author = {Lovelace, R. V. E. and Romanova, M. M. and Bisnovatyi‐Kogan, G. S.},
    number = {1},
    month = {3},
    pages = {368--372},
    volume = {514},
    url = {https://iopscience.iop.org/article/10.1086/306945},
    doi = {10.1086/306945},
    issn = {0004-637X}
}

@article{Manchester2005,
	doi = {10.1086/428488},
	url = {https://doi.org/10.1086/428488},
	year = 2005,
	month = {apr},
	publisher = {{IOP} Publishing},
	volume = {129},
	number = {4},
	pages = {1993--2006},
	author = {R. N. Manchester and G. B. Hobbs and A. Teoh and M. Hobbs},
	title = {The Australia Telescope National Facility Pulsar Catalogue},
	journal = {\aj}
}

@ARTICLE{Mc2006,
   author = {{McLaughlin}, M.~A. and {Lyne}, A.~G. and {Lorimer}, D.~R. and 
	{Kramer}, M. and {Faulkner}, A.~J. and {Manchester}, R.~N. and 
	{Cordes}, J.~M. and {Camilo}, F. and {Possenti}, A. and {Stairs}, I.~H. and 
	{Hobbs}, G. and {D'Amico}, N. and {Burgay}, M. and {O'Brien}, J.~T.
	},
    title = "{Transient radio bursts from rotating neutron stars}",
  journal = {\nat},
   eprint = {astro-ph/0511587},
     year = 2006,
    month = feb,
   volume = 439,
    pages = {817-820},
      doi = {10.1038/nature04440},
   adsurl = {http://adsabs.harvard.edu/abs/2006Natur.439..817M}
}

@article{Mereghetti2021,
   author = {S. Mereghetti and M. Rigoselli and R. Taverna and L. Baldeschi and S. Crestan and R. Turolla and S. Zane},
   doi = {10.3847/1538-4357/AC34F2},
   issn = {0004-637X},
   issue = {2},
   journal = {\apj},
   month = {12},
   pages = {253},
   publisher = {IOP Publishing},
   title = {NICER Study of Pulsed Thermal X-Rays from Calvera: A Neutron Star Born in the Galactic Halo?},
   volume = {922},
   url = {https://iopscience.iop.org/article/10.3847/1538-4357/ac34f2 https://iopscience.iop.org/article/10.3847/1538-4357/ac34f2/meta},
   year = {2021}
}

@ARTICLE{Michel1988,
       author = {{Michel}, F. Curtis},
        title = "{Neutron star disk formation from supernova fall-back and possible observational consequences}",
      journal = {\nat},
         year = 1988,
        month = jun,
       volume = {333},
       number = {6174},
        pages = {644-645},
          doi = {10.1038/333644a0},
       adsurl = {https://ui.adsabs.harvard.edu/abs/1988Natur.333..644M}
}

@ARTICLE{Page+2006,
       author = {{Page}, Dany and {Geppert}, Ulrich and {Weber}, Fridolin},
        title = "{The cooling of compact stars}",
      journal = {\nphysa},
         year = 2006,
        month = oct,
       volume = {777},
        pages = {497-530},
          doi = {10.1016/j.nuclphysa.2005.09.019},
archivePrefix = {arXiv},
       eprint = {astro-ph/0508056},
 primaryClass = {astro-ph},
       adsurl = {https://ui.adsabs.harvard.edu/abs/2006NuPhA.777..497P}
}

@ARTICLE{Page+2009,
       author = {{Page}, Dany and {Lattimer}, James M. and {Prakash}, Madappa and {Steiner}, Andrew W.},
        title = "{Neutrino Emission from Cooper Pairs and Minimal Cooling of Neutron Stars}",
      journal = {\apj},
         year = 2009,
        month = dec,
       volume = {707},
       number = {2},
        pages = {1131-1140},
          doi = {10.1088/0004-637X/707/2/1131},
archivePrefix = {arXiv},
       eprint = {0906.1621},
 primaryClass = {astro-ph.SR},
       adsurl = {https://ui.adsabs.harvard.edu/abs/2009ApJ...707.1131P}
}

@ARTICLE{Perna2000,
       author = {{Perna}, Rosalba and {Hernquist}, Lars and {Narayan}, Ramesh},
        title = "{Emission Spectra of Fallback Disks around Young Neutron Stars}",
      journal = {\apj},
         year = 2000,
        month = sep,
       volume = {541},
       number = {1},
        pages = {344-350},
          doi = {10.1086/309404},
archivePrefix = {arXiv},
       eprint = {astro-ph/9912297},
 primaryClass = {astro-ph},
       adsurl = {https://ui.adsabs.harvard.edu/abs/2000ApJ...541..344P}
}

@ARTICLE{Perna2014,
       author = {{Perna}, Rosalba and {Duffell}, Paul and {Cantiello}, Matteo and
         {MacFadyen}, Andrew I.},
        title = "{The Fate of Fallback Matter around Newly Born Compact Objects}",
      journal = {\apj},
         year = 2014,
        month = feb,
       volume = {781},
       number = {2},
          eid = {119},
        pages = {119},
          doi = {10.1088/0004-637X/781/2/119},
archivePrefix = {arXiv},
       eprint = {1312.4981},
 primaryClass = {astro-ph.HE},
       adsurl = {https://ui.adsabs.harvard.edu/abs/2014ApJ...781..119P}
}

@article{Posselt2018,
	doi = {10.3847/1538-4357/aad6df},
	url = {https://doi.org/10.3847/1538-4357/aad6df},
	year = 2018,
	month = {sep},
	publisher = {\apj},
	volume = {865},
	number = {1},
	pages = {1},
	author = {B. Posselt and G. G. Pavlov and {\"U}. Ertan and {\c{S}}. {\c{C}}al{\i}{\c{s}}kan and K. L. Luhman and C. C. Williams},
	title = {Discovery of Extended Infrared Emission around the Neutron Star {RXJ}0806.4{\textendash}4123},
	journal = {\apj}
}

@ARTICLE{Pons2009,
       author = {{Pons}, J.~A. and {Miralles}, J.~A. and {Geppert}, U.},
        title = "{Magneto-thermal evolution of neutron stars}",
      journal = {\aap},
         year = 2009,
        month = mar,
       volume = {496},
       number = {1},
        pages = {207-216},
          doi = {10.1051/0004-6361:200811229},
archivePrefix = {arXiv},
       eprint = {0812.3018},
 primaryClass = {astro-ph},
       adsurl = {https://ui.adsabs.harvard.edu/abs/2009A&A...496..207P}
}

@article{Rutledge2008,
   author = {R. E. Rutledge and D. B. Fox and A. H. Shevchuk},
   doi = {10.1086/522667/FULLTEXT/},
   issn = {0004-637X},
   issue = {2},
   journal = {\apj},
   month = {1},
   pages = {1137-1143},
   publisher = {American Astronomical Society},
   title = {Discovery of an Isolated Compact Object at High Galactic Latitude},
   volume = {672},
   url = {https://iopscience.iop.org/article/10.1086/522667 https://iopscience.iop.org/article/10.1086/522667/meta},
   year = {2008}
}

@article{Rigoselli2024,
   author = {M. Rigoselli and S. Mereghetti and J. P. Halpern and E. V. Gotthelf and C. G. Bassa},
   doi = {10.3847/1538-4357/AD8CD6},
   issn = {0004-637X},
   issue = {2},
   journal = {\apj},
   month = {11},
   pages = {228},
   publisher = {IOP Publishing},
   title = {The Proper Motion of the High Galactic Latitude Pulsar Calvera},
   volume = {976},
   url = {https://iopscience.iop.org/article/10.3847/1538-4357/ad8cd6 https://iopscience.iop.org/article/10.3847/1538-4357/ad8cd6/meta},
   year = {2024}
}

@ARTICLE{Shakura1973,
   author = {{Shakura}, N.~I. and {Sunyaev}, R.~A.},
    title = "{Black holes in binary systems. Observational appearance.}",
  journal = {\aap},
     year = 1973,
   volume = 24,
    pages = {337-355},
   adsurl = {http://adsabs.harvard.edu/abs/1973A%26A....24..337S}
}

@article{Shevchuk2009,
   author = {Andrew S. H. Shevchuk and Derek B. Fox and Robert E. Rutledge},
   doi = {10.1088/0004-637X/705/1/391},
   issue = {1},
   journal = {ApJ},
   month = {7},
   pages = {391-397},
   publisher = {Institute of Physics Publishing},
   title = {Chandra Observations of 1RXS J141256.0+792204 (Calvera)},
   volume = {705},
   url = {http://arxiv.org/abs/0907.4352 http://dx.doi.org/10.1088/0004-637X/705/1/391},
   year = {2009}
}

@ARTICLE{Trumper2010,
   author = {{Tr{\"u}mper}, J.~E. and {Zezas}, A. and {Ertan}, {\"U}. and 
	{Kylafis}, N.~D.},
    title = "{The energy spectrum of anomalous X-ray pulsars and soft gamma-ray repeaters}",
  journal = {\aap},
archivePrefix = "arXiv",
   eprint = {1004.3391},
 primaryClass = "astro-ph.HE",
     year = 2010,
    month = jul,
   volume = 518,
      eid = {A46},
    pages = {A46},
      doi = {10.1051/0004-6361/200911834},
   adsurl = {http://adsabs.harvard.edu/abs/2010A%26A...518A..46T}
}

@ARTICLE{Trumper2013,
   author = {{Tr{\"u}mper}, J.~E. and {Dennerl}, K. and {Kylafis}, N.~D. and 
	{Ertan}, {\"U}. and {Zezas}, A.},
    title = "{An Accretion Model for the Anomalous X-Ray Pulsar 4U 0142+61}",
  journal = {\apj},
archivePrefix = "arXiv",
   eprint = {1212.5373},
 primaryClass = "astro-ph.HE",
     year = 2013,
    month = feb,
   volume = 764,
      eid = {49},
    pages = {49},
      doi = {10.1088/0004-637X/764/1/49},
   adsurl = {http://adsabs.harvard.edu/abs/2013ApJ...764...49T}
}

@ARTICLE{Ustyugova2006,
       author = {{Ustyugova}, G.~V. and {Koldoba}, A.~V. and {Romanova}, M.~M. and
         {Lovelace}, R.~V.~E.},
        title = "{``Propeller'' Regime of Disk Accretion to Rapidly Rotating Stars}",
      journal = {\apj},
         year = "2006",
        month = "Jul",
       volume = {646},
       number = {1},
        pages = {304-318},
          doi = {10.1086/503379},
archivePrefix = {arXiv},
       eprint = {astro-ph/0603249},
 primaryClass = {astro-ph},
       adsurl = {https://ui.adsabs.harvard.edu/abs/2006ApJ...646..304U}
}

@article{Zane2011,
   author = {S. Zane and F. Haberl and G. L. Israel and A. Pellizzoni and M. Burgay and R. P. Mignani and R. Turolla and A. Possenti and P. Esposito and D. Champion and R. P. Eatough and E. Barr and M. Kramer},
   doi = {10.1111/J.1365-2966.2010.17619.X},
   issn = {0035-8711},
   issue = {4},
   journal = {MNRAS},
   month = {2},
   pages = {2428-2445},
   publisher = {Oxford Academic},
   title = {Discovery of 59 ms pulsations from 1RXS J141256.0+792204 (Calvera)},
   volume = {410},
   url = {https://dx.doi.org/10.1111/j.1365-2966.2010.17619.x},
   year = {2011}
}








\bsp	
\label{lastpage}
\end{document}